\documentclass[conference]{IEEEtran} 
\IEEEoverridecommandlockouts
\usepackage{cite}
\usepackage{amsmath,amssymb,amsfonts}
\usepackage{algorithmic}
\usepackage{graphicx}
\usepackage{textcomp}
\usepackage{xcolor}
\def\BibTeX{{\rm B\kern-.05em{\sc i\kern-.025em b}\kern-.08em
    T\kern-.1667em\lower.7ex\hbox{E}\kern-.125emX}}
\begin{document}

\title{Hypersparse Neural Network Analysis of \\ Large-Scale Internet Traffic
\thanks{This material is based in part upon work supported by the NSF under grants DMS-1312831, CCF-1533644, and CNS-1513283, DHS cooperative agreement FA8750-18-2-0049, and ASD(R\&E) under contract FA8702-15-D-0001.  Any opinions, findings, and conclusions or recommendations expressed in this material are those of the authors and do not necessarily reflect the views of the NSF, DHS, or ASD(R\&E).}
}

\author{\IEEEauthorblockN{Jeremy Kepner$^{1}$, Kenjiro Cho$^{2}$, KC Claffy$^{3}$, Vijay Gadepally$^{1}$, Peter Michaleas$^{1}$, Lauren Milechin$^{4}$
\\
\IEEEauthorblockA{$^1$MIT Lincoln Laboratory Supercomputing Center, $^2$Research Laboratory, Internet Initiative Japan, Inc., \\ $^3$UCSD Center for Applied Internet Data Analysis, $^4$MIT Dept. of Earth, Atmospheric, \& Planetary Sciences
}}}

\maketitle

\begin{abstract}
The Internet is transforming our society, necessitating a quantitative understanding of Internet traffic.  Our team collects and curates the largest publicly available Internet traffic data containing 50 billion packets. Utilizing a novel hypersparse neural network analysis of ``video'' streams of this traffic using 10,000 processors in the MIT SuperCloud reveals a new phenomena: the importance of otherwise unseen leaf nodes and isolated links in Internet traffic.  Our neural network approach further shows that a two-parameter modified Zipf-Mandelbrot distribution accurately describes a wide variety of source/destination statistics on moving sample windows ranging from 100{,}000 to 100{,}000{,}000 packets over collections that span years and continents. The inferred model parameters distinguish different network streams and the model leaf parameter strongly correlates with the fraction of the traffic in different underlying network topologies.  The hypersparse neural network pipeline is highly adaptable and different network statistics and training models can be incorporated with simple changes to the image filter functions.
\end{abstract}


\section{Introduction}
Our civilization is now dependent on the Internet, necessitating a scientific understanding of this virtual universe \cite{hilbert2011world,li2013survey}, that is made more urgent by the rising influence of adversarial Internet robots (botnets) on society \cite{allcott2017social,BadBotReport}.  The two largest efforts to capture, curate, and share Internet packet traffic data for scientific analysis are led by our team via the Widely Integrated Distributed Environment (WIDE) project \cite{cho2000tr} and the Center for Applied Internet Data Analysis (CAIDA) \cite{claffy1999internet}.  These data have supported a variety of research projects resulting in hundreds of peer-reviewed publications \cite{CAIDApubs}, ranging from characterizing the global state of Internet traffic, to specific studies of the prevalence of peer-to-peer filesharing, to testing prototype software designed to stop the spread of Internet worms.


\begin{figure}
\includegraphics[width=1.0\columnwidth]{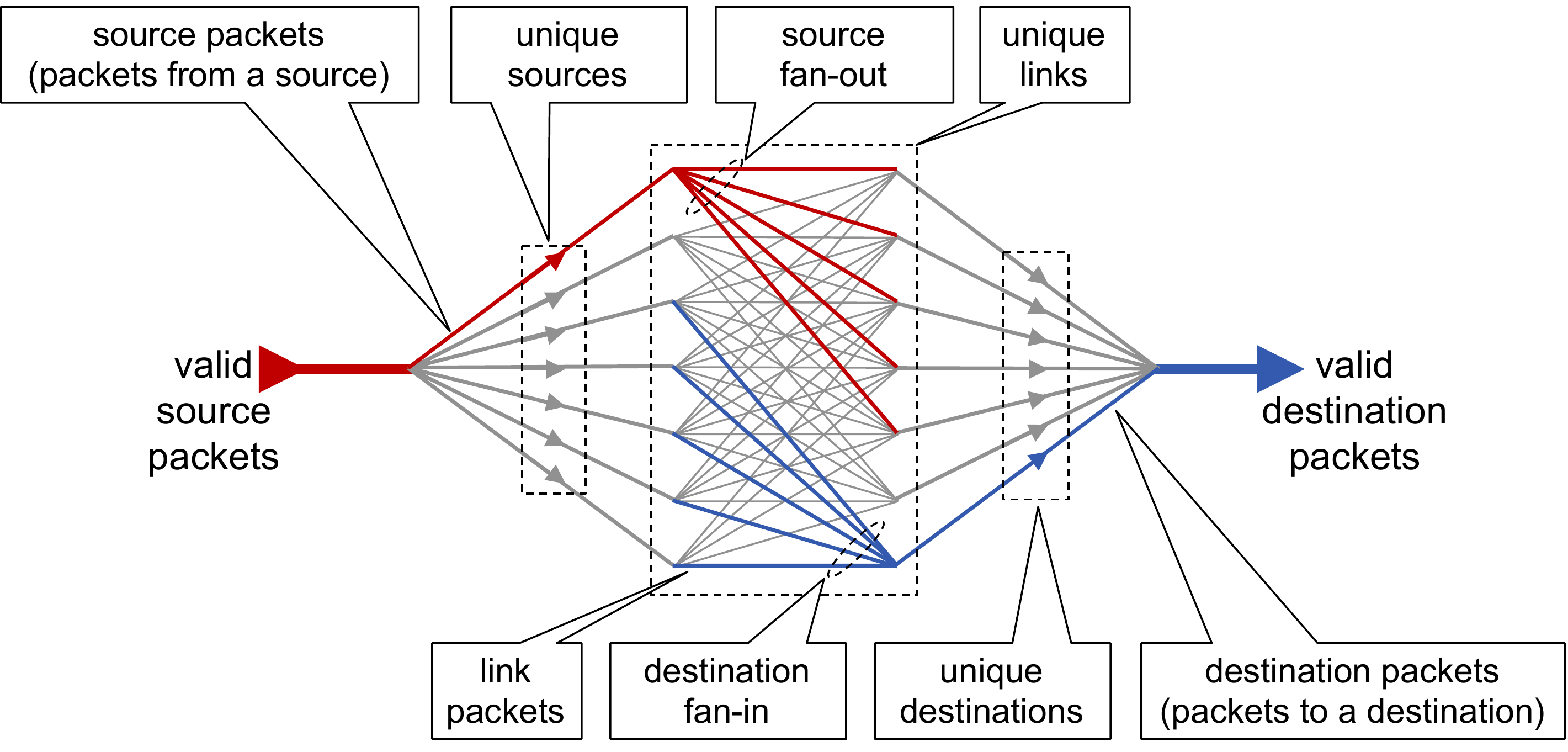}
      	\caption{{\bf Streaming network traffic quantities.} Internet traffic streams of $N_V$ valid packets are divided into a variety of quantities for analysis: source packets, source fan-out, unique source-destination pair packets (or links), destination fan-in, and destination packets.}
      	\label{fig:NetworkDistribution}
\end{figure}

\begin{figure}
\includegraphics[width=1.0\columnwidth]{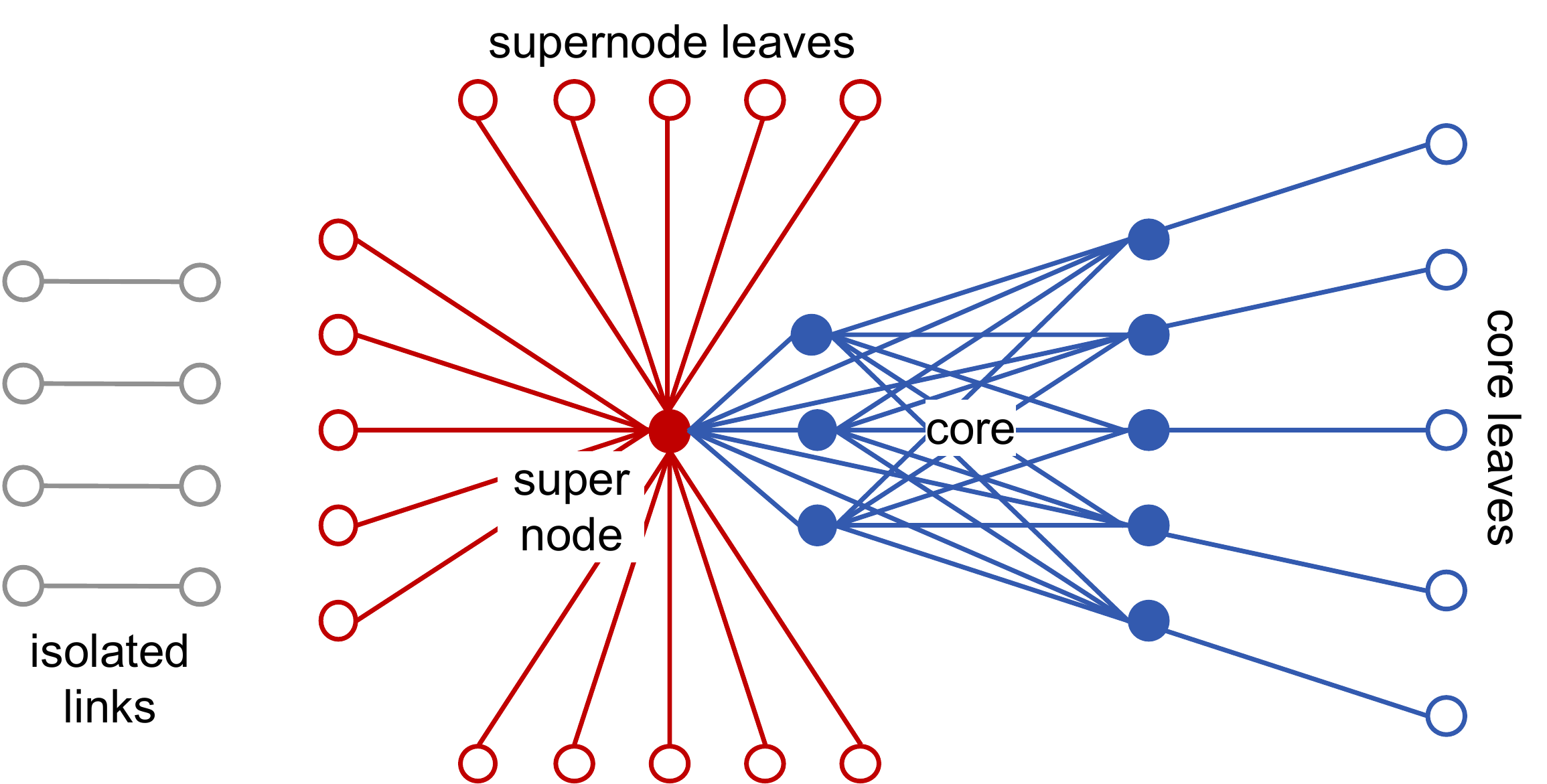}
      	\caption{{\bf Traffic network topologies.} Internet traffic forms networks consisting of a variety of topologies: isolated links, supernode leaves connected to a supernode, densely connected core(s) with corresponding core leaves.}
      	\label{fig:NetworkTopology}
\end{figure}


The stochastic network structure of Internet traffic is a core property of great interest to Internet stakeholders \cite{li2013survey} and network scientists \cite{barabasi2016network}.  Of particular interest is the probability distribution $p(d)$ where $d$ is the degree (or count) of one of several network quantities  depicted in Figure~\ref{fig:NetworkDistribution}: source packets, source fan-out, packets over a unique source-destination pair (or link), destination fan-in, and destination packets.
Amongst the earliest and most widely cited results of virtual Internet topology analysis has been the observation of the power-law relationship
\begin{equation}
 p(d) \propto 1/d^\alpha
\end{equation}
with a model exponent $1 < \alpha < 3$ for large values of $d$ \cite{barabasi1999emergence,albert1999internet,leskovec2005graphs}.
[Note: in our work network topology refers to the graph theoretic virtual topology of sources and destinations and not the underlying physical topology of the Internet.] 
These early observations demonstrated the importance of a few supernodes in the Internet (see Figure~\ref{fig:NetworkTopology})\cite{cao2009identifying}.  Measurements of power-laws in Internet data stimulated investigations into a wide range of network phenomena in many domains and lay the foundation for the field of network science \cite{barabasi2016network}.

Classification of Internet phenomena is often based on data obtained from crawling the network from a number of starting points \cite{olston2010web}.  These webcrawls naturally sample the supernodes of the network \cite{cao2009identifying} and their resulting $p(d)$ are accurately fit at large values of $d$ by single-parameter power-law models.  Unfortunately, these models have impractically large deviations for other values of $d$ (see \cite{clauset2009power} figures 8H/9W/9X, \cite{mahanti2013tale} figure 4B, \cite{kitsak2015long} figure 3A, and  
\cite{lischke2016analyzing} figure 21) and are not usable for modeling Internet traffic in real-world settings.
Characterizing a network by a single power-law exponent provides one view of Internet phenomena, but more accurate and complex models are required to understand the diverse topologies seen in streaming samples of the Internet.

Improving Internet model accuracy while also increasing model complexity requires overcoming a number of challenges, including acquisition of larger, rigorously collected data sets \cite{soule2004identify,zhang2005estimating}; the enormous computational cost of processing large network traffic graphs \cite{lumsdaine2007challenges,bader2013graph,tune2013internet}; careful filtering, binning, and normalization of the data; and inferring nonlinear models to the data \cite{barabasi2016network,clauset2009power}. This paper presents approaches for overcoming these challenges to improved model accuracy by employing a novel hypersparse neural network analysis of ``video'' stream representations of Internet traffic.   Furthermore, the hypersparse neural network pipeline is highly adaptable and different network statistics and training models can be incorporated with simple changes to the image filter functions.

\section{Streaming Internet Data}

The two largest efforts to capture, curate, and share Internet packet traffic data for scientific analysis are led by our team via the WIDE and CAIDA efforts.  This paper analyzes, for the first time, the very largest collections in our corpora containing 49.6 billion packets (see Table~\ref{tab:TrafficData}).

\subsection{MAWI Internet Traffic Collection}
  
  The WIDE project is a research consortium in Japan established in 1988 \cite{cho2000tr}. The members of the project include network engineers, researchers, university students, and industrial partners. The focus of WIDE is on the empirical study of the large-scale internet. WIDE operates an internet testbed for both commercial traffic and for conducting research experiments.  These data have enabled quantitative analysis of Internet traffic spanning years illustrating trends such as, the emergence of residential usage, peer-to-peer networks, probe scanning, and botnets \cite{cho2006impact,borgnat2009seven,fontugne2017scaling}. The Tokyo datasets are publicly available packet traces provided by the WIDE project (a.k.a. the MAWI traces).  The traces are collected from a 1~Gbps academic backbone connection in Japan.  The 2015 and 2017 datasets are 48-hour-long traces captured during December 2-3 2015 and April 12-13 2017 in JST. The IP addresses appearing in the traces are anonymized using a prefix-preserving method \cite{fan2004prefix}.

\begin{table}[htp]
\caption{Packet capture data.}
\vspace{-0.25cm}
Large-scale network traffic packet data sets containing 49.6 billion packets collected at different locations, times, and durations over two years. All source data can be found at the websites https://mawi.wide.ad.jp (/mawi/ditl/ditl2015/ and /mawi/ditl/ditl2017/) and https://www.caida.org (/datasets/passive-2016/equinix-chicago/). This work used the CAIDA UCSD Anonymized Internet Traces - 2016 January 21, February 18,  March 17, and April 06.
\begin{center}
\begin{tabular}{lcccc}
\hline
{\bf Location} & {\bf Date} & {\bf Duration} & {\bf Bandwidth} & {\bf Packets} \\
\hline
Tokyo     & 2015 Dec 02 & 2 days &   $10^9$~ bits/sec &   $17.0{\times}10^9$ \\ 
Tokyo     & 2017 Apr 12 & 2 days &   $10^9$~ bits/sec &   $16.8{\times}10^9$ \\ 
Chicago A & 2016 Jan 21 & 1 hour & $10^{10}$ bits/sec & ~~$2.0{\times}10^9$  \\ 
Chicago A & 2016 Feb 18 & 1 hour & $10^{10}$ bits/sec & ~~$2.0{\times}10^9$  \\ 
Chicago A & 2016 Mar 17 & 1 hour & $10^{10}$ bits/sec & ~~$1.8{\times}10^9$  \\ 
Chicago A & 2016 Apr 06 & 1 hour & $10^{10}$ bits/sec & ~~$1.8{\times}10^9$  \\ 
Chicago B & 2016 Jan 21 & 1 hour & $10^{10}$ bits/sec & ~~$2.3{\times}10^9$  \\ 
Chicago B & 2016 Feb 18 & 1 hour & $10^{10}$ bits/sec & ~~$1.7{\times}10^9$  \\ 
Chicago B & 2016 Mar 17 & 1 hour & $10^{10}$ bits/sec & ~~$2.0{\times}10^9$  \\ 
Chicago B & 2016 Apr 06 & 1 hour & $10^{10}$ bits/sec & ~~$2.1{\times}10^9$  \\ 
\hline
\end{tabular}
\end{center}
\label{tab:TrafficData}
\end{table}%

\begin{figure*}
\includegraphics[width=2.0\columnwidth]{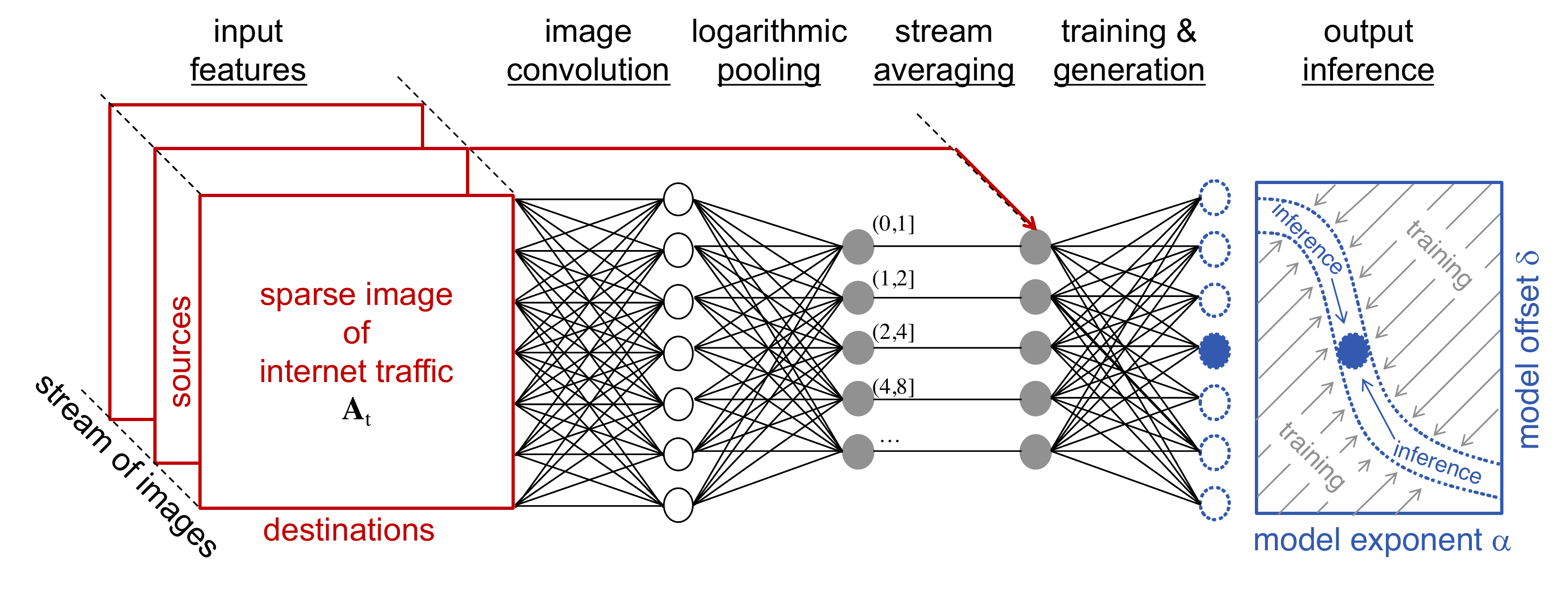}
      	\caption{{\bf Hypersparse neural network pipeline.} Traffic streams are turned into hypersparse images (stored as associative arrays).  Network quantities are extracted from the images via convolution with different filters.  The resulting network quantities are logarithmically pooled (binned) and averaged over the streaming ``video'' of the images.  The averaged data is used to train and then generate network model weights that are used to infer the classification of the network within the model parameter space.}
      	\label{fig:NeuralNetPipeline}
\end{figure*}

\subsection{CAIDA Internet Traffic Collection}

 CAIDA collects several different data types at geographically and topologically diverse locations, and makes this data available to the research community to the extent possible while preserving the privacy of individuals and organizations who donate data or network access \cite{claffy1999internet}\cite{claffy2000measuring}. CAIDA has (and had) monitoring locations in Internet Service Providers (ISPs) in the United States.   CAIDA's passive traces dataset contains traces collected from high-speed monitors on a commercial backbone link. The data collection started in April 2008 and is ongoing. These data are useful for research on the characteristics of Internet traffic, including application breakdown (based on TCP/IP ports), security events, geographic and topological distribution, flow volume and duration. For an overview of all traces see the trace statistics page \cite{CAIDAstats}. Collectively, our consortium has enabled scientific analysis of Internet traffic resulting in hundreds of peer-reviewed publications with over 30,000 citations \cite{CAIDApubs}.

The traffic traces used in this paper  are anonymized using CryptoPan prefix-preserving anonymization. The anonymization key changes annually and is the same for all traces recorded during the same calendar year. During capture packets are truncated at a snap length selected to avoid excessive packet loss due to disk I/O overload. The snap length has historically varied from 64 to 96 bytes. In addition, payload is removed from all packets: only header information up to layer 4 (transport layer) remains.  Endace network cards used to record these traces provide timestamps with nanosecond precision. However, the anonymized traces are stored in pcap format with timestamps truncated to microseconds. Starting with the 2010 traces the original nanosecond timestamps are provided as separate ascii files alongside the packet capture files.

\section{Approach}

This work overcomes obstacles to improved model accuracy by employing a novel hypersparse neural network analysis of ``video'' stream representations of the Internet traffic (Figure~\ref{fig:NeuralNetPipeline}).    Utilizing recent innovations in interactive supercomputing \cite{Kepner2009,reuther2018interactive}, matrix-based graph theory \cite{kolda2009tensor,kepner2011graph}, and big data mathematics  \cite{kepner2018mathematics}, we have developed a scalable neural network Internet traffic processing pipeline that runs efficiently on more than 10{,}000 processors in the MIT SuperCloud \cite{gadepally2018hyperscaling}.  This neural network pipeline allows us, for the first time, to process our largest traffic collections as network traffic graphs.

The hypersparse neural network pipeline depicted in Figure~\ref{fig:NeuralNetPipeline} begins with the construction of sparse images of network traffic data.  These images are then convolved with a filter corresponding to the specific network quantity being analyzed: source packets, source fan-out, links, destination fan-in, and destination packets.  The resulting network quantities are logarithmically pooled (binned) and averaged over the streaming ``video'' of the images.  The averaged data is used to train and then generate network model weights that are used to infer the classification of the network within the model parameter space.

\begin{table}
\caption{Aggregate Network Properties}
\vspace{-0.25cm}
Formulas for computing aggregates from a sparse network image ${\bf A}_t$ at time $t$ in both summation and matrix notation. ${\bf 1}$ is a column vector of all 1's, $^{\sf T}$  is the transpose operation, and $|~|_0$ is the zero-norm that sets each nonzero value of its argument to 1\cite{karvanen2003measuring}.
\begin{center}
\begin{tabular}{p{1in}p{1in}p{0.5in}}
\hline
{\bf Aggregate} & {\bf Summation} & {\bf ~Matrix} \\
{\bf Property} & {\bf ~~Notation} & {\bf Notation} \\
\hline
Valid packets $N_V$ & $\sum_i ~ \sum_j ~ {\bf A}_t(i,j)$ & $~{\bf 1}^{\sf T} {\bf A}_t {\bf 1}$ \\
Unique links & $\sum_i ~ \sum_j |{\bf A}_t(i,j)|_0$  & ${\bf 1}^{\sf T}|{\bf A}_t|_0 {\bf 1}$ \\
Unique sources & $\sum_i |\sum_j ~ {\bf A}_t(i,j)|_0$  & ${\bf 1}^{\sf T}|{\bf A}_t {\bf 1}|_0$ \\
Unique destinations & $\sum_j |\sum_i ~ {\bf A}_t(i,j)|_0$ & $|{\bf 1}^{\sf T} {\bf A}_t|_0 {\bf 1}$ \\
\hline
\end{tabular}
\end{center}
\label{tab:Aggregates}
\end{table}%
\begin{table}[h]
\caption{Neural Network Image Convolution Filters}
\vspace{-0.25cm}
Different network quantities are extracted from a sparse traffic image ${\bf A}_t$ at time $t$ via convolution with different filters.  Formulas for the filters are given in both summation and matrix notation. ${\bf 1}$ is a column vector of all 1's, $^{\sf T}$  is the transpose operation, and $|~|_0$ is the zero-norm that sets each nonzero value of its argument to 1\cite{karvanen2003measuring}.
\begin{center}
\begin{tabular}{p{1.25in}p{1in}p{0.5in}}
\hline
{\bf Network} & {\bf Summation} & {\bf ~Matrix} \\
{\bf Quantity} & {\bf ~~Notation} & {\bf Notation} \\
\hline
Source packets from $i$ & $\sum_j ~ {\bf A}_t(i,j)$ & ~~~$~{\bf A}_t ~~ {\bf 1}$ \\
Source fan-out from $i$ & $\sum_j |{\bf A}_t(i,j)|_0$  & ~~~$|{\bf A}_t|_0 {\bf 1}$ \\
Link packets from $i$ to $j$ & $~~~~~~{\bf A}_t(i,j)$ & ~~~$~{\bf A}_t$ \\
Destination fan-in to $j$ & $\sum_i |{\bf A}_t(i,j)|_0$ & ${\bf 1}^{\sf T}~{\bf A}_t$ \\
Destination packets to $j$ & $\sum_i ~ {\bf A}_t(i,j)$ & ${\bf 1}^{\sf T}|{\bf A}_t|_0$ \\
\hline
\end{tabular}
\end{center}
\label{tab:Filters}
\end{table}%

\subsection{Image Convolution}

Origin-destination traffic matrices or images are one of the most generally useful representations of Internet traffic \cite{tune2013internet,soule2004identify}.  These matrices can be used to compute a wide range of network statistics useful in the analysis, monitoring, and control of the Internet. Such analysis include the temporal fluctuations of the supernodes \cite{soule2004identify} and inferring the presence of unobserved traffic \cite{zhang2005estimating}\cite{bharti2010inferring}.  Not all packets have both source and destination Internet protocol version 4 (IPv4) addresses.  To reduce statistical fluctuations the streaming data have been partitioned so that for any chosen time window all data sets have the same number of valid IPv4 packets.  At a given time $t$, $N_V$ consecutive valid packets are aggregated from the traffic into a sparse matrix ${\bf A}_t$, where ${\bf A}_t(i,j)$ is the number of valid packets between the source $i$ and destination $j$ \cite{mucha2010community}. The sum of all the entries in ${\bf A}_t$ is equal to $N_V$
  \begin{equation}\label{eq:Valid}
    \sum_{i,j} {\bf A}_t(i,j) = N_V
  \end{equation}
All the network quantities depicted in Figure~\ref{fig:NetworkDistribution} can be readily computed from ${\bf A}_t$ using the formulas listed in Table~\ref{tab:Aggregates} and Table~\ref{tab:Filters}.

\begin{figure}
\includegraphics[width=\columnwidth]{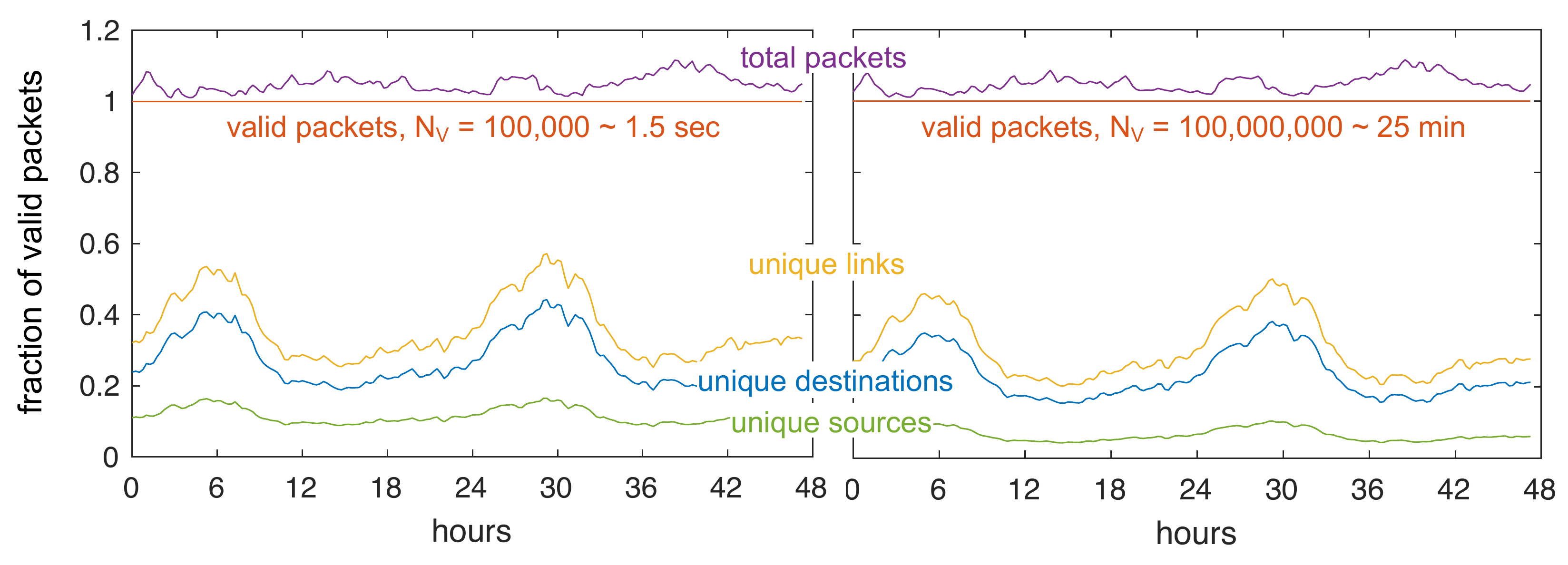}
\includegraphics[width=\columnwidth]{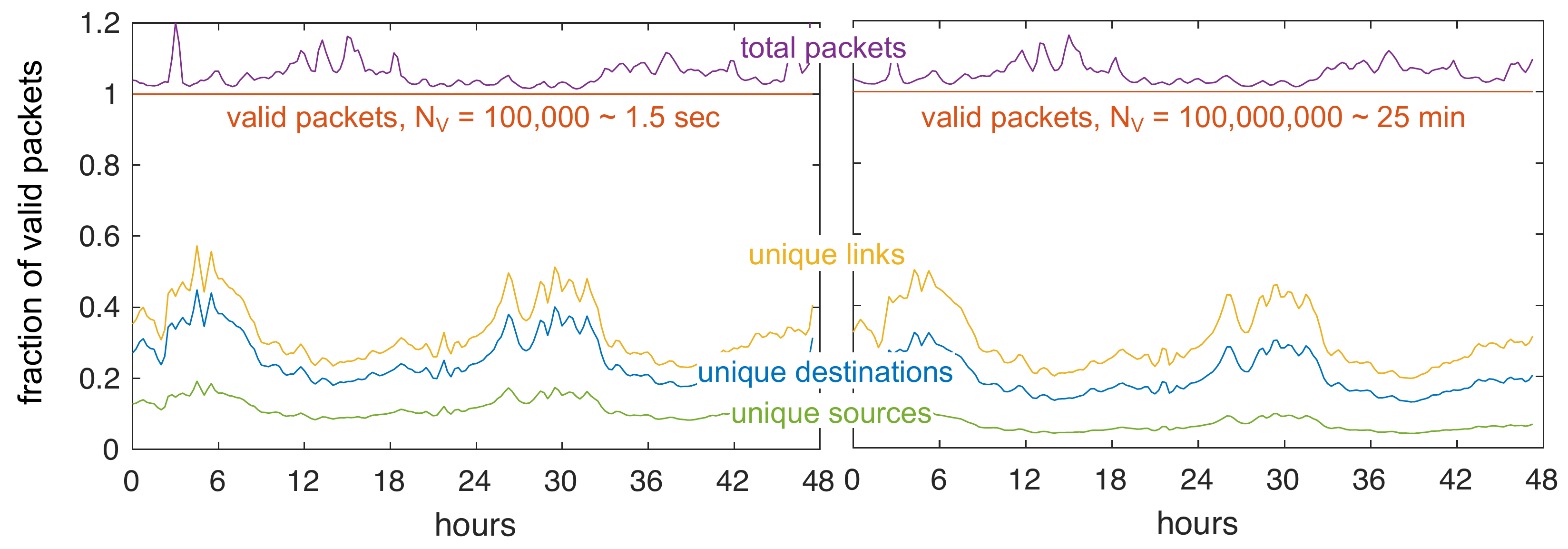}
      	\caption{{\bf Valid packets.} Analyzing packet windows with the same numbers of valid packets produces consistent fractions of the aggregate numbers of unique links, unique destinations, and unique sources over a wide range of packet sizes for the Tokyo 2015 (top) and Tokyo 2017 (bottom) data sets. The plots show these fractions for moving packet windows of with $N_V$ = 100{,}000 packets (left) and $N_V$ = 100{,}000{,}000 packets (right).  The packet windows correspond to time windows of approximately 1.5 seconds and 25 minutes.
	}
      	\label{fig:ValidPackets}
\end{figure}

An essential step for increasing the accuracy of the statistical measures of Internet traffic is using windows with the same number of valid packets $N_V$.  For this analysis, a valid packet is defined as TCP over IPv4, which includes more than 95\% of the data in the collection and eliminates a small amount of data that uses other protocols or contains anomalies.  Using packet windows with the same number of valid packets produces aggregates that are consistent over a wide range from $N_V = 100{,}000$ to $N_V = 100{,}000{,}000$ (Figure~\ref{fig:ValidPackets}).

\subsection{Logarithmic Pooling}

A network quantity $d$ is computed via convolution with the image ${\bf A}_t$ using a filter selected from Table~\ref{tab:Filters}.  The corresponding histogram of the network quantity is denoted by $n_t(d)$, with corresponding probability
  \begin{equation}\label{eq:Probability}
    p_t(d) = n_t(d)/\sum_d n_t(d)
  \end{equation}
and cumulative probability
  \begin{equation}\label{eq:Cumulative}
    P_t(d) = \sum_{i=1,d} p_t(d)
  \end{equation}
Because of the relatively large values of $d$ observed due to a single supernode, the measured probability at large $d$ often exhibits large fluctuations. However, the cumulative probability lacks sufficient detail to see variations around specific values of $d$, so it is typical to pool the differential cumulative probability with logarithmic bins in $d$
  \begin{equation}\label{eq:LogBin}
    D_t(d_i) = P_t(d_i) - P_t(d_{i-1})
  \end{equation}
where $d_i = 2^i$ \cite{clauset2009power}.  All computed probability distributions use the same binary logarithmic pooling (binning) to allow for consistent statistical comparison across data sets (Eq.~\ref{eq:Cumulative})\cite{clauset2009power,barabasi2016network}.  The corresponding mean and standard deviation of $D_t(d_i)$ over many different consecutive values of $t$ for a given data set are denoted $D(d_i)$ and $\sigma(d_i)$.


\subsection{Modified Zipf-Mandelbrot Model}

Measurements of $D(d_i)$ can reveal many properties of network traffic, such as the number of nodes with only one connection $D(d = 1)$ and the size of the supernode
\begin{equation}
  d_{\rm max}={\rm argmax}(D(d) > 0)
\end{equation}
Effective classification of a network with a low parameter model allows these and many other properties to be summarized and computed efficiently.  In the standard Zipf-Mandelbrot model typically used in linguistic contexts, $d$ is a ranking with $d=1$ corresponding to the most popular value \cite{mandelbrot1953informational,montemurro2001beyond,saleh2006modeling}. To accurately classify the network data using the full range of $d$, the Zipf-Mandelbrot model is modified so that $d$ is a measured network quantity instead of a rank index
  \begin{equation}\label{eq:ZipfMandelbrot}
    p(d;\alpha,\delta) \propto 1/(d + \delta)^\alpha
  \end{equation}
The inclusion of a second model offset parameter $\delta$ allows the model to accurately fit small values of $d$, in particular $d=1$, which has the highest observed probability in these streaming data. The model exponent $\alpha$ has a larger impact on the model at large values of $d$ while the model offset $\delta$ has a larger impact on the model at small values of $d$ and in particular at $d=1$.


The unnormalized modified Zipf-Mandelbrot model is denoted
  \begin{equation}\label{eq:rho}
    \rho(d;\alpha,\delta) = \frac{1}{(d + \delta)^\alpha}
  \end{equation}
with correspond gradient
  \begin{equation}\label{eq:drho}
    \partial_\delta \rho(d;\alpha,\delta) = \frac{-\alpha}{(d + \delta)^{\alpha+1}} = -\alpha \rho(d;\alpha+1,\delta)
  \end{equation}
The normalized model probability is given by
  \begin{equation}\label{eq:ZM}
    p(d;\alpha,\delta) = \frac{\rho(d;\alpha,\delta)}{\sum_{d=1}^{d_{\rm max}} \rho(d;\alpha,\delta)}
  \end{equation}
where $d_{max}$ is the largest value of the network quantity $d$.  The cumulative model probability is the sum 
  \begin{equation}\label{eq:ZMcum}
    P(d_i;\alpha,\delta) = \sum_{d=1}^{d_i} p(d;\alpha,\delta)
  \end{equation}
The corresponding differential cumulative model probability is
  \begin{equation}\label{eq:ZMdiff}
    D(d_i;\alpha,\delta) = P(d_i;\alpha,\delta) - P(d_{i-1};\alpha,\delta)
  \end{equation}
where $d_i = 2^i$.

\begin{figure*}
\centerline{\includegraphics[width=2\columnwidth]{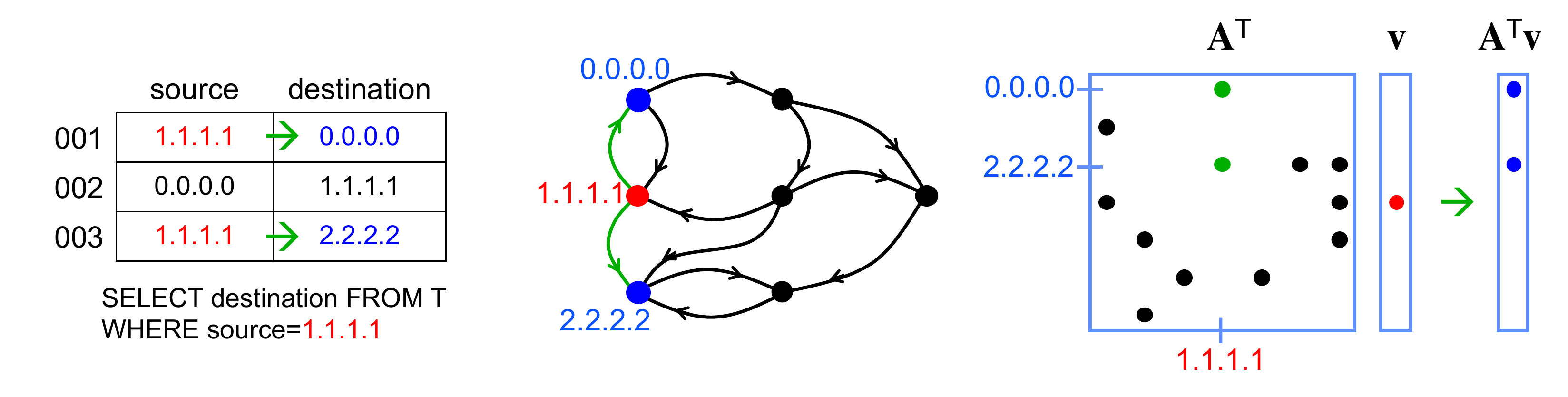}}
      	\caption{{\bf Associative Arrays.} Hypersparse network data are naturally represented as associative  arrays that uniquely label each row and and column.  Associative arrays further allow the data to be readily manipulated using relational database operations, graph algorithms, and matrix mathematics. (left) Tabular representation of raw network traffic and corresponding database query to find all records beginning with source 1.1.1.1. (middle) Network graph highlighting nearest neighbors of source node 1.1.1.1.  (right) Corresponding associative array representation of the network graph illustrating how the neighbors of source node 1.1.1.1 are computed with matrix vector multiplication.}
      	\label{fig:AssociativeArrays}
\end{figure*}

\subsection{Training and Weight Generation}


Classifying the logarithmically pooled data in terms of model parameters $\alpha$ and $\delta$ begins with training a set of candidate model weights that can then be used to infer the model parameters. Initially, a set of candidate exponent values is selected, typically $\alpha = 0.10, 0.11,\ldots,3.99,4.00$. For each value of $\alpha$, a value of $\delta$ is trained that exactly matches the model with the data at $D(1)$.  Training the value of $\delta$ corresponding to a give $D(1)$ is done using the gradient based Newton's method as follows.  Setting the measured value of $D(1)$ equal to the model value $D(1;\alpha,\delta)$ gives
  \begin{equation}\label{eq:ZM1}
    D(1) = D(1;\alpha,\delta) = \frac{1}{(1 + \delta)^{\alpha} \sum_{d=1}^{d_{\rm max}} \rho(d;\alpha,\delta)}
  \end{equation}
Newton's method works on functions of the form $f(\delta) = 0$. Rewriting the above expression produces
  \begin{equation}\label{eq:ZMnewton}
    f(\delta) = D(1) (1 + \delta)^\alpha \sum_{d=1}^{d_{\rm max}} \rho(d;\alpha,\delta) - 1 = 0
  \end{equation}
For given value of $\alpha$, $\delta$ can be trained using the following iterative gradient based equation
  \begin{equation}\label{eq:NewtonIteration}
     \delta \rightarrow \delta - \frac{f(\delta)}{\partial_\delta f(\delta)}
  \end{equation}
where the gradient is

$\partial_\delta f(\delta) = $
  \begin{eqnarray}\label{eq:NewtonDerivative}
       && \alpha D(1) (1 + \delta)^\alpha  \\
       && \Big[(1 + \delta)^{-1} \sum_{d=1}^{d_{\rm max}} \rho(d;\alpha,\delta) - 
                   \sum_{d=1}^{d_{\rm max}} \rho(d;\alpha+1,\delta)\Big] \nonumber
  \end{eqnarray}

Using a starting value of $\delta=1$ and bounds of $0 < \delta < 10$, Newton's method can be iterated until the differences in successive values of $\delta$ fall below a specified error (typically 0.001) and is usually achieved in less than five iterations.


\subsection{Parameter Inference}

The inferred $\alpha$ (and corresponding $\delta$) is chosen by minimizing the $|~|^{1/2}$ metric over logarithmic differences between the weights of candidate models $D(d_i;\alpha,\delta)$ and the data 
  \begin{equation}\label{eq:NonLinFit}
    {\rm argmin}_{\alpha} \sum_{d_i}|\log(D(d_i)) - \log(D(d_i;\alpha,\delta))|^{1/2}
  \end{equation}
The $|~|^{1/2}$ metric (or $|~|_p$-norm with $p = 1/2$) favors maximizing error sparsity over minimizing outliers\cite{donoho2006compressed,chartrand2007exact,xu2012}\cite{karvanen2003measuring,saito2000sparsity,Brbic2018,Rahimi2018scale}. Several authors have shown recently that it is possible to reconstruct a nearly sparse signal from fewer linear measurements than would be expected from traditional sampling theory.  Furthermore, by replacing the $|~|_1$ norm with the $|~|^p$ with $p < 1$,  reconstruction is possible with substantially fewer measurements.

Using logarithmic values more evenly weights their contribution to the inferred model and more accurately reflects the number of packets used to compute each value of $D(d_i)$.  Lower accuracy data points are avoided by limiting the training and inference procedure to data points where the value is greater than the standard deviation: $D(d_i) > \sigma(d_i)$.

\subsection{Memory and Computation Requirements}

Processing 49.6 billion Internet packets with a variety of algorithms presents numerous computational challenges.  Dividing the data set into combinable units of approximately 100{,}000 consecutive packets made the analysis amenable to processing on a massively parallel supercomputer.  The detailed architecture of the parallel processing system and its corresponding performance are described in \cite{gadepally2018hyperscaling}.  The resulting processing pipeline was able to efficiently use over 10{,}000 processors on the MIT SuperCloud and was essential to this first-ever complete analysis of these data.

A key element of our analysis is the use of novel hypersparse matrix mathematics in concert with the MIT SuperCloud to process very large network traffic matrices (Figure~\ref{fig:AssociativeArrays}).  Construction and analysis of network traffic matrices of the entire Internet address space have been considered impractical for its massive size\cite{tune2013internet}.  Internet Protocol version 4 (IPv4) has $2^{32}$ unique addresses, but at any given collection point only a fraction of these addresses will be observed.  Exploiting this property to save memory can be accomplished by extending traditional sparse matrices so that new rows and columns can be added dynamically.  The algebra of associative arrays \cite{kepner2018mathematics} and its corresponding implementation in the Dynamic Distributed Dimensional Data Model (D4M) software library (d4m.mit.edu) allows the row and columns of a sparse matrix to be any sortable value, in this case character string representations of the Internet addresses (Figure~\ref{fig:AssociativeArrays}).  Associative arrays extend sparse matrices to have database table properties with dynamically insertable and removable rows and columns that adjust as new data is added or subtracted to the matrix.  Using these properties, the memory requirements of forming network traffic matrices can be reduced at the cost of increasing the required computation necessary to resort the rows and columns.


A hypersparse associative array representing an image of traffic ${\bf A}_t$ with $N_V = 100{,}000{,}000$ typically requires 2 Gigabytes of memory.   Complete analysis of the statistics and topologies of ${\bf A}_t$ typically takes 10 minutes on a single MIT SuperCloud Intel Knights Landing processor core.   Using increments of $100{,}000$ packets means that this analysis is repeated over 500{,}000 times to process all 49.6 billion packets.  Using 10{,}000 processors on the MIT SuperCloud shortens the run time of these analysis to approximately eight hours.  The results presented here are the product of an interactive discovery process that required hundreds of such runs that would not have been possible without the MIT SuperCloud.  Fortunately, the utilization of these results by Internet stakeholders can be significantly accelerated by creating optimized embedded hypersparse neural network implementations that only compute the desired statistics and are not required to support an interactive analytics discovery process \cite{liu2010tcam,liu2016packet}.

\section{Results}

\subsection{Daily Variations}

\begin{figure}
\centering
\includegraphics[width=1.0\columnwidth]{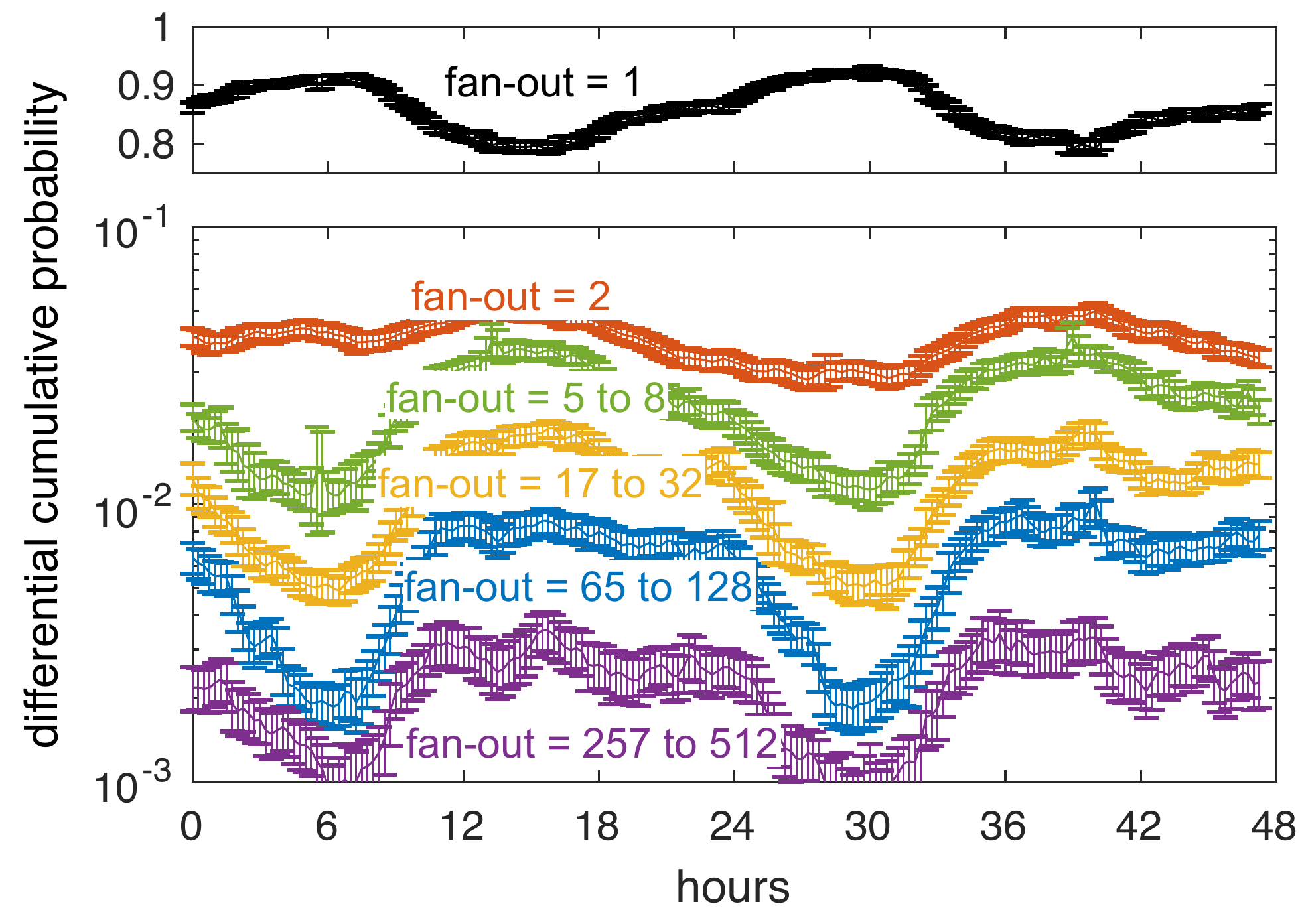}
      	\caption{{\bf Daily variation in Internet traffic.}  The fraction of source nodes with a given range of fan-out are shown as a function of time for the Tokyo 2015 data.  The $p(d = 1)$ value is plotted on a separate linear scale because of the larger magnitude relative to the other points.  Each point is the mean of many neighboring points in time and the error bars are the measured $\pm$1-$\sigma$.  The daily variation of the distributions oscillate between extremes corresponding to approximately local noon and midnight.
	}
      	\label{fig:DailyVariation}
\end{figure}

\begin{figure}
\centering
\includegraphics[width=1.0\columnwidth]{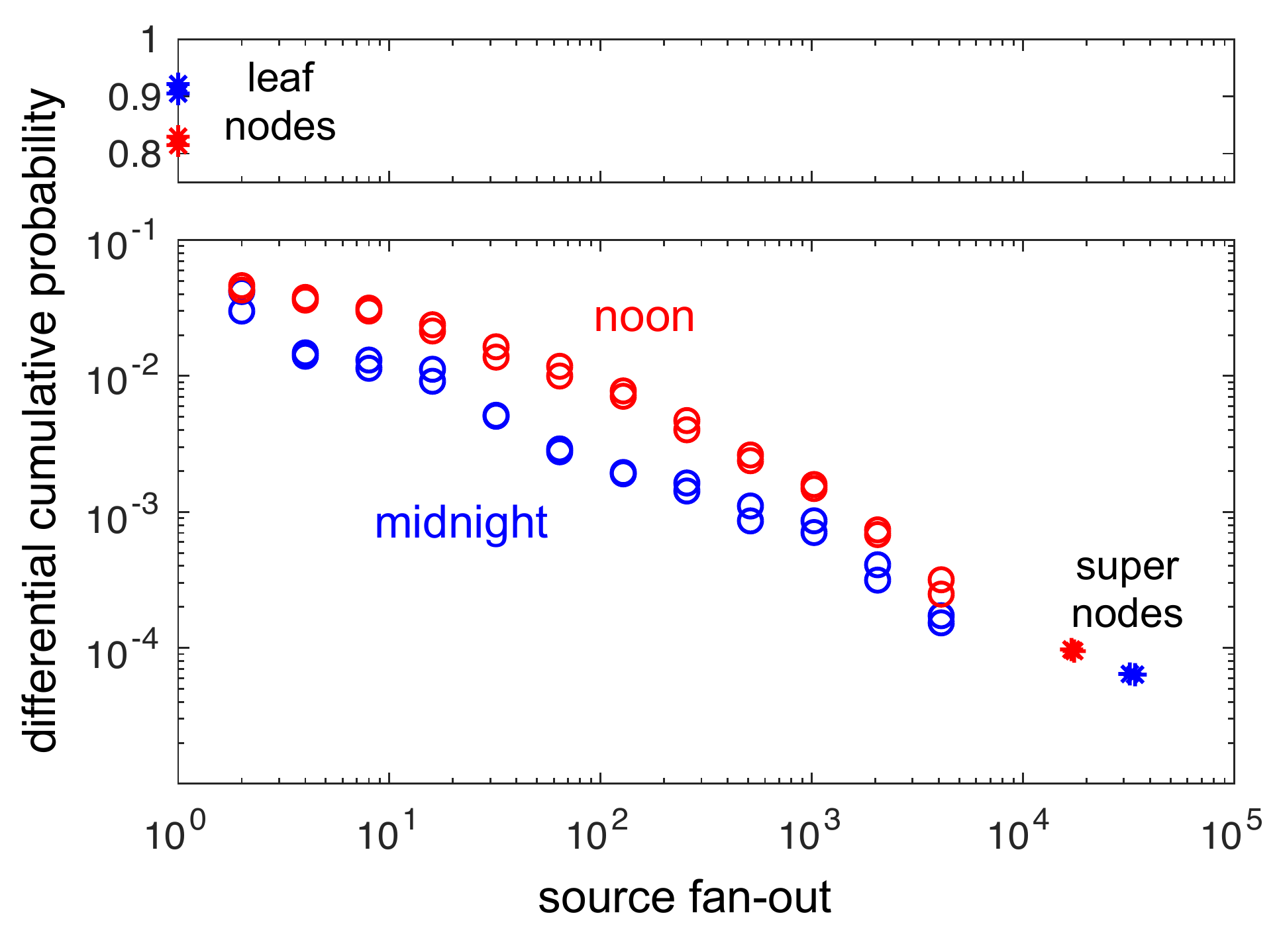}
      	\caption{{\bf Daily limits in Internet traffic.}  The fraction of source nodes versus fan-out are shown for two noons and two midnights for the Tokyo 2015 data.  The overlap among the noons and the midnights shows the relative day-to-day consistency in these data and show the limits of the two extremes in daily variation.  During the day, there is more traffic among nodes with intermediate fan-out.  At night the traffic is more dominated by leaf nodes and the supernode.
	}
      	\label{fig:DailyLimits}
\end{figure}

Diurnal variations in supernode network traffic are well known \cite{soule2004identify}. The Tokyo packet data were collected over a period spanning two days, and allow the daily variations in packet traffic to be observed.  The precision and accuracy of our measurements allows these variations to be observed across a wide range of nodes. Figure~\ref{fig:DailyVariation} shows the fraction of source fan-outs in each of various bin ranges.  The fluctuations show the network evolving between two envelopes occurring  between noon and midnight that are shown in Figure~\ref{fig:DailyLimits}.  

\subsection{Inferred Modified Zipf-Mandelbrot Distributions}

\begin{figure*}
\includegraphics[width=2.0\columnwidth]{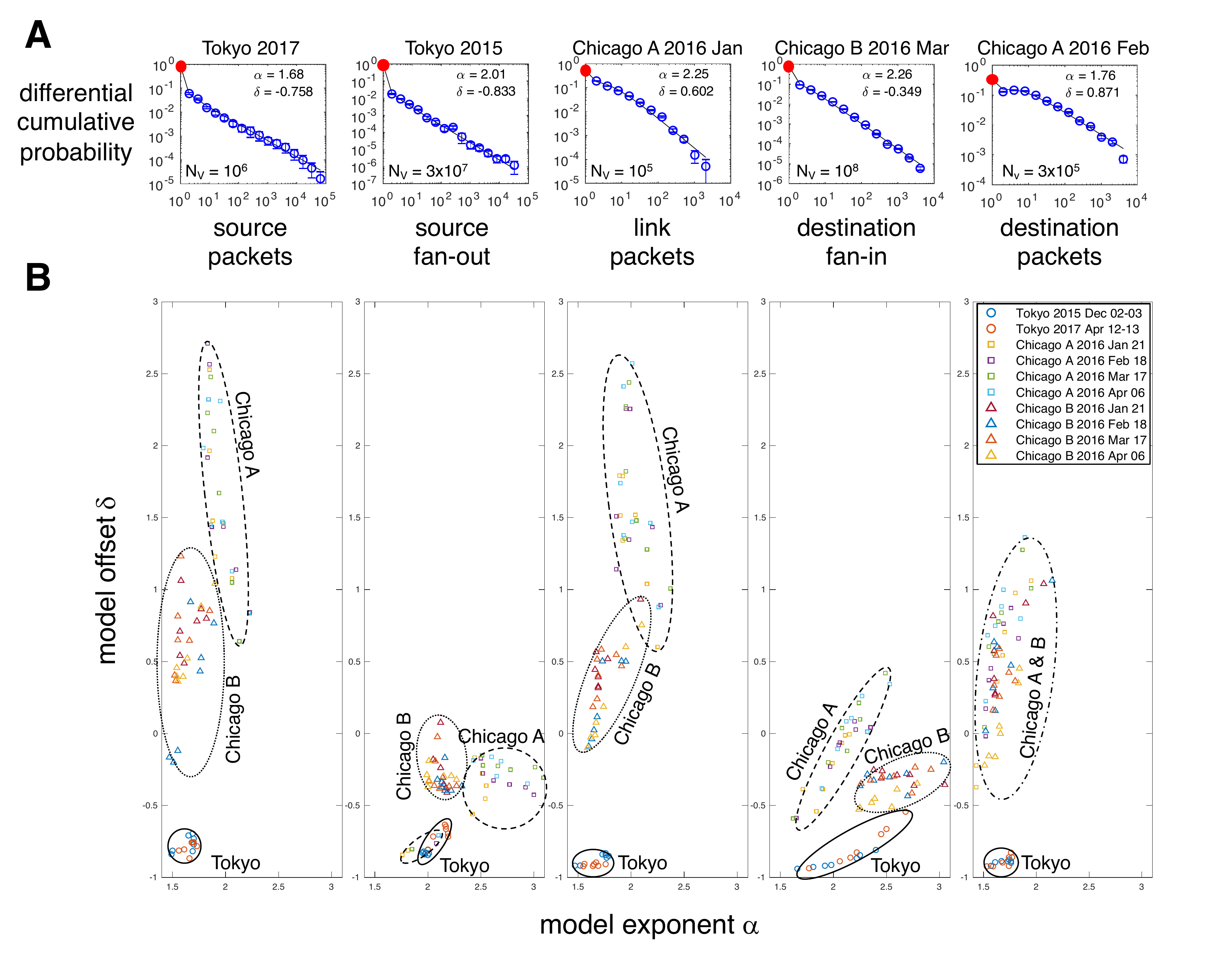}
      	\caption{{\bf Measured network traffic distributions and inferred models.} ({\bf{\sf A}}) A selection of 5 of the 350  measured differential cumulative probabilities
spanning different locations, dates, and packet windows.  Blue circles are measured data with $\pm$1-$\sigma$ error bars.  Black lines are the best-fit modified Zipf-Mandelbrot models with parameters $\alpha$ and $\delta$.  Red dots highlight the large contribution of leaf nodes and isolated links.  ({\bf{\sf B}}) Inferred model parameters for all 350 measured probability distributions reveal the underlying structural differences among the data collected in Tokyo, Chicago A, and Chicago B.}
      	\label{fig:DistributionParameters}
\end{figure*}

Figure~\ref{fig:DistributionParameters}A
shows five representative inferred models out of the 350 performed on 10 datasets, 5 network quantities, and 7 valid packet windows: $N_V = 10^5$, $3{\times}10^5$, $10^6$, $3{\times}10^6$, $10^7$, $3{\times}10^7$, $10^8$.
The inferred models are valid over the entire range of $d$ and provide parameter estimates with precisions of 0.01.  In every case, the high value of $p(d=1)$ is indicative of a large contribution from a combination of supernode leaves, core leaves, and isolated links (Figure~\ref{fig:NetworkTopology}). The breadth and accuracy of these data allow detailed comparison of the inferred models.  Figure~\ref{fig:NetworkDistribution}B shows the inferred model offset $\delta$ versus the model exponent $\alpha$ for all 350 fits. The different collection locations are clearly distinguishable in this model parameter space.  The Tokyo collections have smaller offsets and are more tightly clustered than the Chicago collections.  Chicago B has a consistently smaller source and link packet model offset than Chicago A.  All the collections have source, link, and destination packet model exponents in the relatively narrow $1.5 < \alpha < 2$ range.  The source fan-out and destination fan-in model exponents are in the broader $1.5 < \alpha < 2.5$ range and are consistent with the prior literature \cite{clauset2009power}.  These results represent an entirely new approach to characterizing Internet traffic that allows the data to be projected into a low-dimensional space and enables accurate comparisons among packet collections with different locations, dates, durations, and sizes.

\subsection{Measured Network Topologies}

\begin{figure*}
\includegraphics[width=2.0\columnwidth]{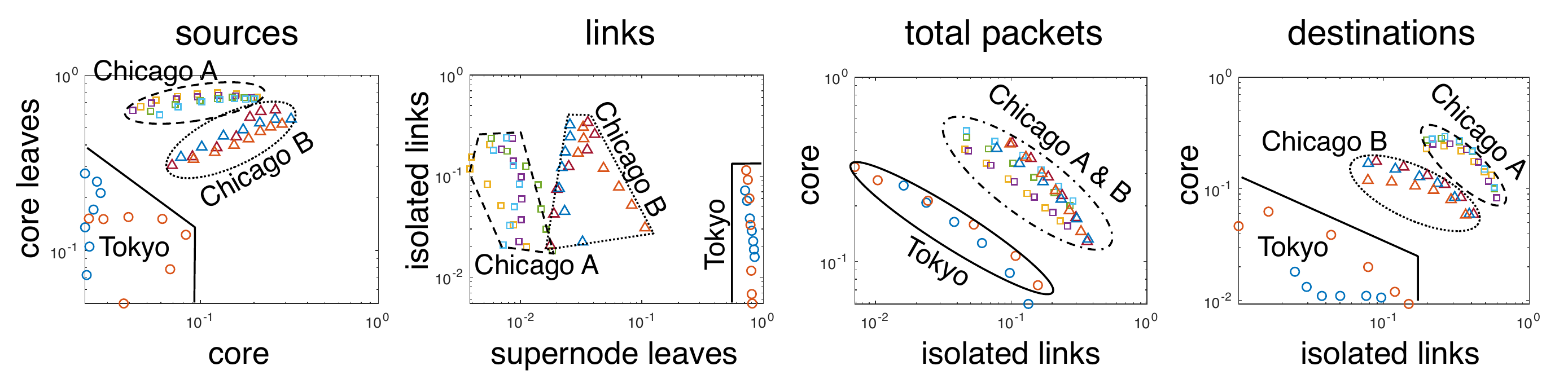}
      	\caption{{\bf Distribution of traffic among network topologies.}  A selection of four projections
showing the fraction of data in various underlying topologies using the same legend as Figure~\ref{fig:DistributionParameters}B.  Horizontal and vertical axis are the corresponding fraction of the sources, links, total packets and destinations that are in various topologies for each location, time, and seven packet windows ($N_V = 10^5, \ldots, 10^8$). These data reveal the differences in the network traffic topologies in the data collected in Tokyo (dominated by supernode leaves), Chicago A (dominated by core leaves), and Chicago B (between Tokyo and Chicago A).}
      	\label{fig:MeasuredTopology}
\end{figure*}

\begin{figure*}
\includegraphics[width=2.0\columnwidth]{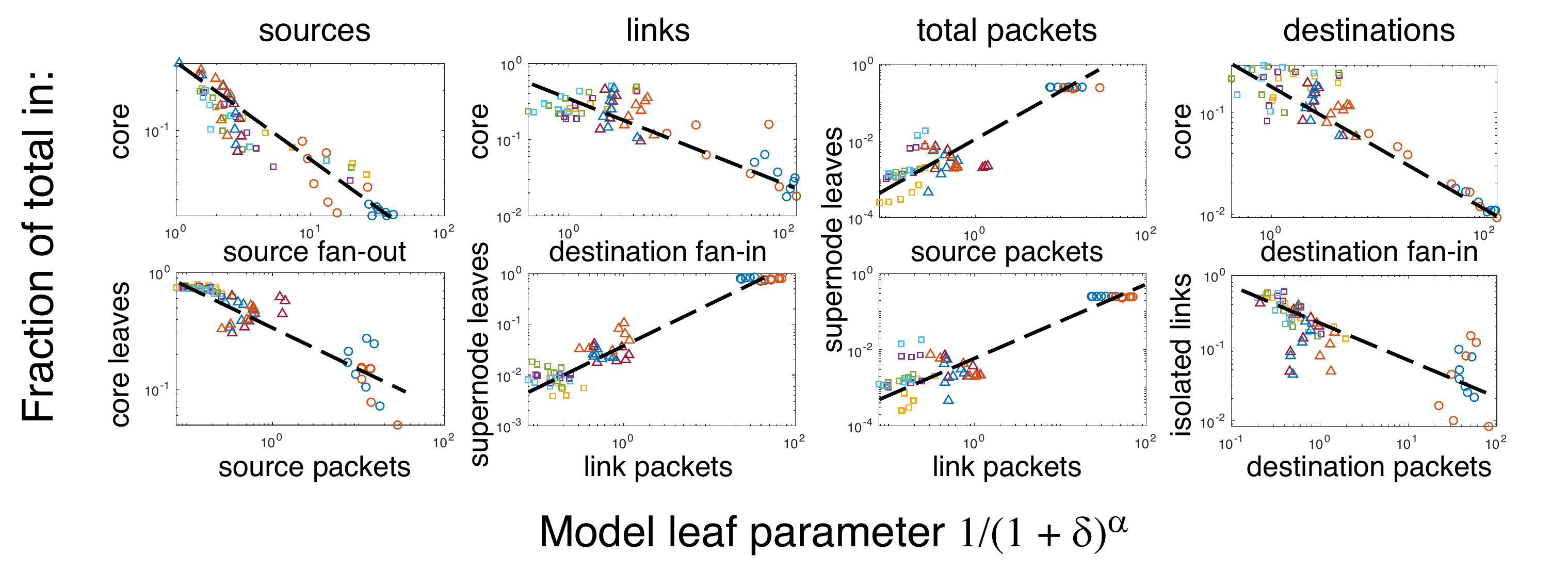}
      	\caption{{\bf Topology versus model leaf parameter.} Network topology is highly correlated with the modified Zipf-Mandelbrot model leaf parameter $1/(1+\delta)^\alpha$. A selection of eight projections
showing the fraction of sources, links, total packets, and destinations in various underlying topologies using the same legend as Figure~\ref{fig:DistributionParameters}B.  Vertical axis are the corresponding fraction of the sources, links, total packets and destinations that are in various topologies. Horizontal axis is the value of the model parameter taken from  either the  source packet, source fan-out, link packet, destination fan-in and destination packet fits.  Data points are for each location, time, and seven packet windows ($N_V = 10^5, \ldots, 10^8$).}
      	\label{fig:TopoModel}
\end{figure*}

Figure~\ref{fig:NetworkTopology} depicts the major topological structures in the network traffic: isolated links, supernode leaves, core, and core leaves.  Formulas for computing these topologies from ${\bf A}_t$ are given in Appendix~\ref{sec:TopologyMeasures}.
Figure~\ref{fig:MeasuredTopology} shows the average relative fractions of sources, total packets, total links, and number of destinations in each of the five topologies for the ten data sets, and seven valid packet windows: $N_V = 10^5$, $3{\times}10^5$, $10^6$, $3{\times}10^6$, $10^7$, $3{\times}10^7$, $10^8$.  The four projections in Figure~\ref{fig:MeasuredTopology} were chosen
to highlight the differences in the collection locations.  The distinct  regions in the various projections shown in Figure~\ref{fig:MeasuredTopology} indicate that underlying topological differences are present in the data.  The Tokyo collections have much larger supernode leaf components than the Chicago collections.  The Chicago collections have much larger core and core leaves components than the Tokyo collections.  Chicago A consistently has fewer isolated links than Chicago B.  Comparing the modified Zipf-Mandelbrot model parameters in Figure~\ref{fig:DistributionParameters}B and underlying topologies in Figure~\ref{fig:MeasuredTopology} suggests that the inferred model parameters are a more compact way to distinguish the network traffic.

Figures~\ref{fig:DistributionParameters}B and \ref{fig:MeasuredTopology} indicate that different collection points  produce different inferred model parameters $\alpha$ and $\delta$, and that these collection points also have different underlying topologies. Figure~\ref{fig:TopoModel} connects the inferred models and topology observations by plotting the topology fraction as a function of the model leaf parameter $1/(1+\delta)^\alpha$ which corresponds to the  relative strength of leaf nodes and isolated links $p(d=1)$
  \begin{equation}\label{eq:ZipfMandelbrot}
    1/(1 + \delta)^\alpha \propto p(d=1;\alpha,\delta)
  \end{equation}
The correlations revealed in Figure~\ref{fig:TopoModel} suggest that the model leaf parameter strongly correlates with the fraction of the traffic in different underlying network topologies and is a potentially new and beneficial way to characterize networks.  Figure~\ref{fig:TopoModel} indicates that the fraction of sources, links, and destinations in the core shrinks as the relative importance of the leaf parameter in the source fan-out and destination fan-in increases.  In other words, more source and destination leaves means a smaller core.    Likewise, the fraction of links and total packets in the supernode leaves grows as the leaf parameter in the link packets and source packets increases.  Interestingly, the fraction of sources in the core leaves and isolated links decreases as the leaf parameter in the source and destination packets increases indicating a shift of sources away from the core leaves and isolated links into supernode leaves.  Thus, the modified Zipf-Mandelbrot model and its leaf parameter provide a direct connection with the network topology, underscoring the value of having accurate model fits across the entire range of values and in particular for $d=1$.

\section{Conclusion}

Measurements of internet traffic are useful for informing policy, identifying and preventing outages, defeating attacks, planning for future loads, and protecting the domain name system \cite{clark20179th}. On a given day, millions of IPs are engaged in scanning behavior. Our improved models can aid cybersecurity analysts in determining which of these IPs are nefarious \cite{yu2012predicted}, the distribution of attacks in particular critical sectors \cite{husak2018assessing}, identifying spamming behavior \cite{fonseca2016measuring}, how to vacinate against computer viruses \cite{balthrop2004technological}, obscuring web sources \cite{javed2015measurement}, identifying significant flow aggregates in traffic \cite{cho2017recursive}, and sources of rumors \cite{paluch2018fast}.

The results presented here have a number of potential practical applications for Internet stakeholders.  The methods presented of collecting, filtering, computing, and binning the data to produce accurate measurements of a variety of network quantities are generally applicable to Internet measurement and have the potential to produce more accurate measures of these quantities.  The accurate fits of the two parameter modified Zipf-Mandelbrot distribution offer all the usual benefits of low parameter models: measuring parameters with far less data, accurate predictions of network quantities based on a few parameters, observing changes in the underlying distribution, and using modeled distributions to detect anomalies in the data.  

From a scientific perspective, improved knowledge of how Internet traffic flows can inform our understanding of how economics, topology, and demand shape the Internet over time.  As with all scientific disciplines, the ability of theoreticians to develop and test theories of the Internet and network phenomena is bounded by the scale and accuracy of measured phenomena \cite{adamic2000power,bohman2009emergence,stumpf2012critical,virkar2014power}.
In contrast to previous network models that have principally been based on data obtained from network crawls from a variety of start points on the network, our network traffic data are collected from observations of network streams.  Both viewpoints provide important network observations. Observations of a network stream provide complementary data on network dynamics and highlight the contribution of leaves and isolated edges, which are less sampled in network crawls.

The aggregated data set our teams have collected provide a unique window into these questions.  The hypersparse neural network pipeline is a novel approach for inferring power-law models and have potential applications to power-law networks in diverse domains. The inferred model parameters present new opportunities to connect the distributions to underlying theoretical models of networks. That the inferred model parameters distinguish the different collection points and are reflective of different network topologies in the data at these points suggests a deeper underlying connection between the models and the network topologies.

\section*{Acknowledgment}

The authors wish to acknowledge the following individuals for their contributions and support: Shohei  Araki, William Arcand, David Bestor, William Bergeron, Bob Bond, Paul Burkhardt, Chansup Byun, Cary Conrad, Alan Edelman, Sterling Foster, Bo Hu, Matthew Hubbell, Micheal Houle, Micheal Jones, Anne Klein, Charles Leiserson, Dave Martinez, Mimi McClure, Julie Mullen, Steve Pritchard, Andrew Prout, Albert Reuther, Antonio Rosa, Victor Roytburd, Siddharth Samsi, Koichi Suzuki, Kenji Takahashi, Michael Wright, Charles Yee, and Michitoshi  Yoshida.

\bibliographystyle{ieeetr}
\bibliography{InternetModelNeuralNetwork}

\appendices
\setcounter{equation}{0}
\renewcommand{\theequation}{\thesection\arabic{equation}}

\section{Network Topology Measures}
\label{sec:TopologyMeasures}

Figure~\ref{fig:NetworkTopology} depicts the major topological structures in the network traffic.  Identification of these topologies and computation of their network statistics can all be obtained from the packet traffic counts aggregated into the sparse matrix ${\bf A}_t$.  Two important network quantities for computing these network topologies are the source fan-out column vector
  \begin{equation}\label{eq:dSourceFanOuts}
    {\bf d}_{\rm out} = |{\bf A}_t|_0 {\bf 1}
  \end{equation}
and the destination fan-in row vector
  \begin{equation}\label{eq:dDestinationFanIns}
    {\bf d}_{\rm in} = {\bf 1}^{\sf T}|{\bf A}_t|_0
  \end{equation}

\subsection{Isolated Links}
Isolated links are sources and destinations that each have only one connection.  The set of sources that send to only one destination are
  \begin{equation}\label{eq:i1}
    i_1 = {\rm arg}({\bf d}_{\rm out} = 1)
  \end{equation}
The set of destinations that receive from only one destination are
  \begin{equation}\label{eq:i1}
    j_1 = {\rm arg}({\bf d}_{\rm in} = 1)
  \end{equation}
The isolated links can be found via
  \begin{equation}\label{eq:IsolatedLinks}
    {\bf A}_t(i_1,j_1)
  \end{equation}
The number of isolated link sources are
  \begin{equation}\label{eq:UniqueSourcesIsolated}
    {\bf 1}^{\sf T}|{\bf A}_t(i_1,j_1) {\bf 1}|_0
  \end{equation}
The number of packets traversing isolated links is given by
   \begin{equation}\label{eq:ValidPacketsIsolated}
    {\bf 1}^{\sf T}{\bf A}_t(i_1,j_1){\bf 1}
  \end{equation}
The number of unique isolated links can be computed from
  \begin{equation}\label{eq:UniqueLinksIsolated}
    {\bf 1}^{\sf T}|{\bf A}_t(i_1,j_1)|_0 {\bf 1}
  \end{equation}
The number of isolated link destinations are
  \begin{equation}\label{eq:UniqueDestinationsIsolated}
    |{\bf 1}^{\sf T} {\bf A}_t(i_1,j_1)|_0 {\bf 1}
  \end{equation}
By definition, the number of isolated sources, the number of isolated links, and the number of isolated destinations are all the same value.

\subsection{Supernodes}
The first, second, third, ... supernode is the source or destination with the first, second, third, ... most links.  The identity of the first supernode is given by
  \begin{equation}\label{eq:SuperNode}
    k_{\rm max} = {\rm argmax}({\bf d}_{\rm out} + {\bf d}_{\rm in})
  \end{equation}
The leaves of a supernode are those sources and destinations whose only connection is to the supernode.
The supernode source leaves can be found via
  \begin{equation}\label{eq:SuperSourceLeaves}
    {\bf A}_t(i_1,k_{\rm max})
  \end{equation}
The supernode destination leaves can be found via
  \begin{equation}\label{eq:SuperDestinationLeaves}
    {\bf A}_t(k_{\rm max},j_1)
  \end{equation}
The number of supernode leaf sources are
  \begin{equation}\label{eq:UniqueSourcesSuper}
    {\bf 1}^{\sf T}|{\bf A}_t(i_1,k_{\rm max}) {\bf 1}|_0
  \end{equation}
The number of packets traversing supernode leaves is given by
   \begin{equation}\label{eq:ValidPacketsSuper}
    {\bf 1}^{\sf T}{\bf A}_t(i_1,k_{\rm max}) + {\bf A}_t(k_{\rm max},j_1){\bf 1}
  \end{equation}
The number of unique supernode leaf links can be computed from
  \begin{equation}\label{eq:UniqueLinksSuper}
    {\bf 1}^{\sf T}|{\bf A}_t(i_1,k_{\rm max})|_0 + |{\bf A}_t(k_{\rm max},j_1)|_0 {\bf 1}
  \end{equation}
The number of supernode leaf destinations are
  \begin{equation}\label{eq:UniqueDestinationsSuper}
    |{\bf 1}^{\sf T} {\bf A}_t(k_{\rm max},j_1)|_0 {\bf 1}
  \end{equation}
Subsequent supernodes can be computed by eliminating the prior supernode and repeating the above calculations.

\subsection{Core}
  The core of a network can be defined in a variety of ways.  In this work, the network core is meant to convey the concept of a collection of sources and destinations that are not isolated and are multiply connected.  The core is defined as the collection of sources and destinations whereby every source and destination has more than one connection.  In this context, the core does not include the first five supernodes although only the first supernode is significant, and whether or not the other supernodes are included has minimal impact on the core calculations for these data.  The set of sources that send to more than one destination, excluding the supernode(s), is
  \begin{equation}\label{eq:icore}
    i_{\rm core} = {\rm arg}(1 < {\bf d}_{\rm out} < {\bf d}_{\rm out}(k_{\rm max}))
  \end{equation}
The set of destinations that receive from more than one source, excluding the supernode(s), is
  \begin{equation}\label{eq:jcore}
    j_{\rm core} = {\rm arg}(1 < {\bf d}_{\rm in} < {\bf d}_{\rm in}(k_{\rm max}))
  \end{equation}
The core links can be found via
  \begin{equation}\label{eq:IsolatedCore}
    {\bf A}_t(i_{\rm core},j_{\rm core})
  \end{equation}
The number of core sources is
  \begin{equation}\label{eq:UniqueSourcesCore}
    {\bf 1}^{\sf T}|{\bf A}_t(i_{\rm core},j_{\rm core}) {\bf 1}|_0
  \end{equation}
The number of core packets is given by
   \begin{equation}\label{eq:ValidPacketsCore}
    {\bf 1}^{\sf T}{\bf A}_t(i_{\rm core},j_{\rm core}){\bf 1}
  \end{equation}
The number of unique core links can be computed from
  \begin{equation}\label{eq:UniqueLinksCore}
    {\bf 1}^{\sf T}|{\bf A}_t(i_{\rm core},j_{\rm core})|_0 {\bf 1}
  \end{equation}
The number of core destinations is
  \begin{equation}\label{eq:UniqueDestinationsCore}
    |{\bf 1}^{\sf T} {\bf A}_t(i_{\rm core},j_{\rm core})|_0 {\bf 1}
  \end{equation}

\subsection{Core Leaves}
The core leaves are sources and destinations that have one connection to a core source or destination. The core source leaves can be found via
  \begin{equation}\label{eq:CoreSourceLeaves}
    {\bf A}_t(i_1,k_{\rm core})
  \end{equation}
The core destination leaves can be found via
  \begin{equation}\label{eq:CoreDestinationLeaves}
    {\bf A}_t(k_{\rm core},j_1)
  \end{equation}
The number of core leaf sources is
  \begin{equation}\label{eq:UniqueSourcesCoreLeaves}
    {\bf 1}^{\sf T}|{\bf A}_t(i_1,k_{\rm core}) {\bf 1}|_0
  \end{equation}
The number of core leaf packets is given by
   \begin{equation}\label{eq:ValidPacketsCoreLeaves}
    {\bf 1}^{\sf T}{\bf A}_t(i_1,k_{\rm core}) + {\bf A}_t(k_{\rm core},j_1){\bf 1}
  \end{equation}
The number of unique core leaf links can be computed from
  \begin{equation}\label{eq:UniqueLinksCoreLeaves}
    {\bf 1}^{\sf T}|{\bf A}_t(i_1,k_{\rm core})|_0 + |{\bf A}_t(k_{\rm core},j_1)|_0 {\bf 1}
  \end{equation}
The number of core leaf destinations is
  \begin{equation}\label{eq:UniqueDestinationsCoreLeaves}
    |{\bf 1}^{\sf T} {\bf A}_t(k_{\rm core},j_1)|_0 {\bf 1}
  \end{equation}

\end{document}